\documentclass{optica-article}

\journal{opticajournal} 

\articletype{Research Article}

\usepackage{lineno}

\begin{document}

\title{Free space optical link to a tethered balloon for frequency transfer and chronometric geodesy}

\author{Nicolas Maron,\authormark{1,2} Sébastien Fernandez,\authormark{1,2} François-Xavier Esnault,\authormark{2} Thomas Lévèque,\authormark{2} Tepuaonini Muzeau\authormark{3} and Peter Wolf\authormark{1,*}}

\address{\authormark{1}LNE-SYRTE, Observatoire de Paris, Universit\'e PSL, CNRS, Sorbonne Universit\'e, 61 avenue de l'Observatoire, 75014 Paris, France\\
\authormark{2}Centre National d'Etudes Spatiales, 18 avenue Edouard Belin, 31400 Toulouse, France\\
\authormark{3}Institut National des Sciences Appliquées de Toulouse, 31400 Toulouse, France}

\email{\authormark{*}peter.wolf@obspm.fr} 


\begin{abstract*} 
We present the results of an optical link to a corner cube on board a tethered balloon at 300~m altitude including a Tip/Tilt compensation for the balloon tracking. Our experiment measures the carrier phase of a 1542~nm laser, which is the useful signal for frequency comparison of distant clocks. An active phase noise compensation of the carrier is implemented, demonstrating a fractional frequency stability of $8 \times 10^{-19}$ after 16~s averaging, which slightly (factor $\sim$~3) improves on best previous links via an airborne platform. This state-of-the-art result is obtained with a transportable set-up that enables a fast field deployment.
\end{abstract*}


\section{Introduction}
Optical clocks have demonstrated uncertainties in the low $10^{-18}$ region in fractional frequency, after integration times of order $10^3$ seconds \cite{BACON2021,Takamoto2020,Oelker2019}. This makes them prime candidates for applications in navigation, geodesy, and fundamental physics \cite{Wolf2016a}.

Currently, only optical methods using the carrier phase or femtosecond pulses, allow comparing such clocks without degrading their performance. Such methods have been implemented in optical fibre links \cite{Primas1988,Lopez2012,Lisdat2016,Guillou2018} and more recently through free space \cite{Djerroud2010a,Giorgetta2013,Kang2019,Bergeron2019,DixMatthews2021,Shen2022a,Gozzard2022,DixMatthews2023}. Many of the applications mentioned above require intercontinental and/or satellite links, and efforts to work towards space optical links for ultra-stable clock comparisons are on the way \cite{Chiodo2013a,Shen2021,Caldwell2023}.

In the meantime, ground links between clocks over shorter distances are required, not only as demonstrators for space, but also for field measurement of the local geopotential via the relativistic gravitational redshift of clocks, known as chronometric geodesy  \cite{Lion2017,Bondarescu2015a,Denker2017}. This requires not only a high performance clock (a fractional frequency change of $1\times 10^{-18}$ corresponds to a change in orthometric height of $\sim 1$~cm), but also a practical method to compare the frequency of that clock to a reference clock in a fixed location. The latter would be typically delivered at the nearest access to a fibre network, in general at a distance of order $\sim 100$~km.

Thus, using clocks for centimetric mapping of the geopotential at multiple points in regions with difficult access (mountains, coastline, \dots) \cite{Lion2017}, is expected to be one of the first actual application of transportable optical clocks \cite{Takamoto2020,Huang2020} and medium distance ($\sim 100$~km) free-space optical links. Such links will rarely have direct point-to-point unobstructed lines of sight, and work is on the way to establish links through airborne relays using drones \cite{Bergeron2019,DixMatthews2023}. Those experiments have demonstrated frequency stability of $2.5\times10^{-18}$ (in terms of modified Allan deviation, MDEV) for integration times of a few seconds. The work presented in this paper is along the same lines, with similar (marginally better) performance, but uses a balloon, which brings advantages like slower motion and potentially much higher altitude (up to stratospheric) allowing common views over much larger baselines. Additionally balloons or stratospheric platforms persist over much longer times (days or even weeks) allowing fine study of time varying effects e.g. of tidal, hydrological or geological origin.

Free space optical-link technology for clock comparisons has developed rapidly in the wake of optical clocks, from pioneering experiments demonstrating the feasibility, potential, and main limitations \cite{Djerroud2010a}, to near quantum limited performance compatible with ground to geostationary orbit distances \cite{Caldwell2023}. Three geometries have been used: (i) Genuine point to point links (with a fibre link in parallel for performance evaluation) \cite{DixMatthews2021,Shen2022a}; (ii) Partially folded links with separated emitter and receiver closely co-located \cite{Giorgetta2013,Kang2019,Caldwell2023}; (iii) Fully folded links with common emitter and receiver \cite{Gozzard2022}. Strictly speaking only links of type (i) allow a full, loop-hole free, demonstration of the link performance. Indeed, the performance relies on reduction by several orders of magnitude of the main phase noise sources (e.g. atmospheric turbulence, motion of terminals, aperture related effects) in real time or post-analysis. Such reduction requires reciprocity between the noise of the outward and return paths \cite{Robert2016}, and folded links may introduce correlations between the two paths that would not be there in a realistic scenario of point to point clock comparisons.

Nonetheless, folded links still allow studying key aspects and technological performance for optical clock comparisons, in particular when used in conjunction with airborne relays. The corresponding technological demonstration includes tracking of a moving target, effect of link outages, phase noise compensation functionality and performance, transportability and compactness.

All demonstrations to airborne platforms have so far used fully folded architectures (type (iii) above) \cite{Bergeron2019,DixMatthews2023}, and our experiment is no exception. Nevertheless, in contrast to previous work, our demonstration is based on a transportable and quickly field deployable setup, which implements a link to a tethered balloon rather than drones. Thus, our results are an essential milestone for point to point ground to airborne links, and by extension for ground to ground links via an airborne relay for intermediate distances.

Our article is organised as follows. In section \ref{sec:experiment} we describe the experimental set-up with the results shown in section \ref{sec:results}. We also show results from an equivalent (same distance) folded link with a static target on the ground. In section \ref{sec:discussion} we discuss our results, study the main noise sources, and put them into context with respect to previous work. We conclude with an outlook to future work in section \ref{sec:conclusion}.

\section{Experimental scheme} \label{sec:experiment}

\begin{figure}
\centering
  \centering
  \includegraphics[width=\textwidth]{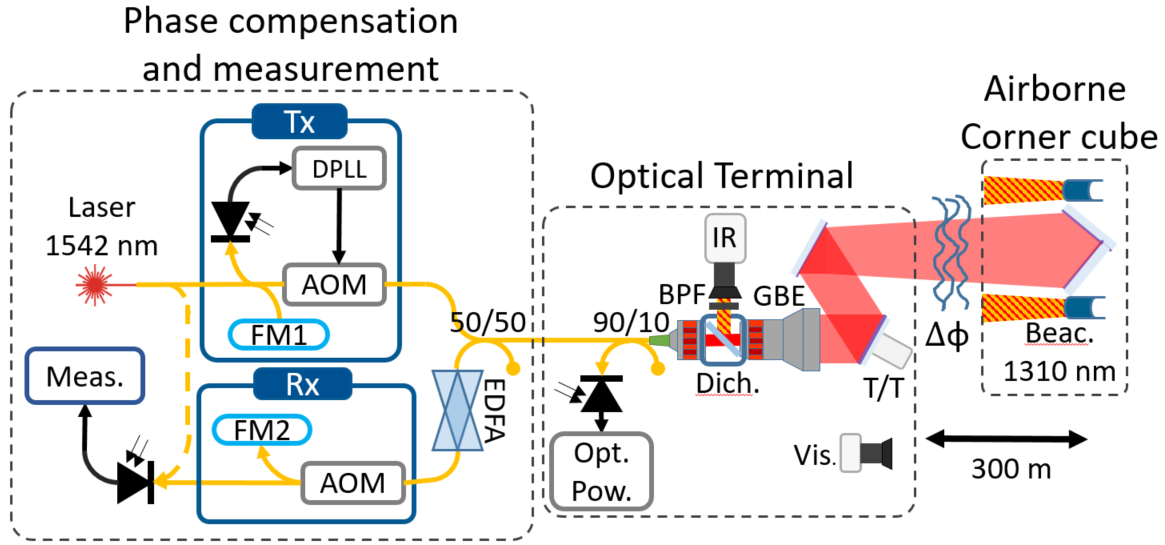}
  \captionof{figure}{Schematic of the experiment}
  \small{
  \textbf{AOM}: Acousto-Optic Modulator;
  \textbf{Beac.}: Beacon laser;
  \textbf{BPF}: 1310 nm Band Pass Filter
  \textbf{Dich.}: Dichroic mirror;
  \textbf{DPLL}: Digital Phase Lock Loop;
  \textbf{EDFA}: Erbium Doped fibre Amplifier (bidirectional);
  \textbf{FM}: Faraday Mirror;
  \textbf{GBE}: Galilean Beam Expander;
  \textbf{IR}: Infrared camera;
  \textbf{Meas.}: out of loop phase Measurement;
  \textbf{Opt. Pow.}: return Optical Power monitoring;
  \textbf{Rx}: Receiver; 
  \textbf{Tx}: Transmitter;
  \textbf{T/T}: Tip/Tilt mirror;
  \textbf{Vis.}: Visible camera;
  \(\Delta \phi\): phase noise from turbulences}
  \label{fig:schematic_exp}
\end{figure}

Our experimental setup, depicted on Fig. \ref{fig:schematic_exp} consists off a free-space phase-compensated laser link between a transportable ground segment and an airborne corner cube. The link stabilization is realized through a heterodyne Michelson architecture and the system allows to perform a simultaneous out-of-loop phase measurement. The ground segment of our system is split into a phase compensation and measurement unit and an optical terminal. The airborne segment consists of a corner cube reflector and four beacon laser diodes attached to a tethered helium balloon. The complete setup was developed and tested at CNES premises in Toulouse and then transported to the Aire sur l'Adour balloon operation centre of CNES for field measurements. The ground segment fits easily into a 6~m$^3$ mini-van and can be deployed by two operators within a short time (typ. $<$~30 minutes between arrival and full lock). Figure \ref{fig:balloon_and_payload} shows images of the ground terminal, payload and balloon during operation.

\subsection{Phase compensation and measurement unit}

The phase compensation and measurement unit enables both to stabilize the phase of the laser link and to perform an out-of-loop measurement of the link stability. To do so, the optical signal is generated from a planar extended cavity 1542~nm laser diode delivering 7~mW into a monomode fibre. The phase noise compensation of the link is based of an unequal-arms Michelson interferometer \cite{Ma1994}. Its short fibre arm ($\sim 1$~m) spanning from the Tx splitter to FM1 (see Fig. \ref{fig:schematic_exp}) serves as a local reference to which the long folded arm (the link) spanning from the Tx splitter to FM2 is compared. The comparison is done by a heterodyne detection on a photodiode at 160~MHz achieved by two AOMs inserted into the long arm. The first AOM (Tx, controlled by the DPLL) shifts the laser frequency by +38~MHz$-\frac{1}{2\pi}\frac{d\Delta\phi(t)}{dt}$ where $2\Delta\phi(t)$ is the round trip phase noise in the long arm seen by the DPLL. The second AOM (Rx) shifts the laser frequency by +42~MHz and enables to discriminate all possible optical path in the folded architecture as well as intermediate parasitic reflections. The 160~MHz beatnote on the Tx photodiode is isolated by a bandpass filter and then divided down to 16~MHz in order to fit in the input range of the DPLL. The DPLL is run on a RedPitaya FPGA board using the software developed in \cite{TourignyPlante2018}. The bandwidth of the full phase-noise compensation loop is $\sim$15~kHz.

The laser source and the receiver being physically close allows for an out of loop phase comparison between the two using small uncompensated fibres (yellow dashed line in Fig. \ref{fig:schematic_exp}). The received laser frequency is shifted by 80~MHz relative to the source. The two are optically beat on a photodiode and generate a 80~MHz RF beatnote. In order for our phasemeter, a Microsemi 53100A, to measure the useful 80~MHz beatnote phase, we must isolate it from neighbouring beatnotes at $\pm$4~MHz generated by alternative optical paths in the folded link. This is achieved by downcoverting the 80~MHz signal (using a mixer) down to 1~MHz and low pass filtering it. The phasemeter and the DPLL are referenced onto a common local oscillator from a frequency generator.
    
\subsection{Optical terminal}

The optical terminal behaves as a transceiver in which the laser light is sent and received through the same optical system. The science signal is received from the phase compensation and measurement unit through a monomode fibre. It is coupled to free space using a f=20~mm fibre collimator followed by a x5 magnification Galilean beam expander (GBE). This optical system generates a Gaussian beam with a waist of $w=9.5$~mm with little clipping from the 35~mm aperture of the GBE. The actuation of the optical terminal is based on a 2" Piezo actuated Tip/Tilt mirror for fast target movements and first order atmospheric turbulence compensation, and an astronomical equatorial mount for high amplitude low frequency target motion tracking.
    
The target tracking relies on four 1310~nm (10° divergence) beacon laser diodes attached to the airborne corner cube and aligned with its optical axis. The beacon lasers are filtered by a dichroic mirror and detected on the terminal by a 215~fps infrared (IR) monochromatic camera. The camera focal length is 75~mm and it uses a 320x256 pixel sensor with 20~$\mu$m pixel size. Taking into account the x5 GBE, this corresponds to a angular pixel size on the sky of 53~$\mu$rad, and to a $\approx$~5x4~m image at the balloon distance of 300~m. The barycenter of each beacon image (spread over $< 9$ pixels) is individually detected in real time using a computer vision algorithm. Their overall centroid position within the image stream is then stabilized by a digital servo loop that controls the Tip/Tilt mirror. The stabilization setpoint is chosen to maximize the science beam optical return into the fibre. Despite the high frame rate, a delay of 16.5~ms between the camera and the control computer was found to be the main limiting factor on our Tip/Tilt correction bandwidth, sitting around 20 Hz. A second servo loop maintains the Tip/Tilt mirror close to the middle of its actuation range (10~mrad) using the astronomical mount. In typical operating conditions lock was never lost during several hours of operation.
  
A wide field-of-view (6°), co-aligned, visible camera allows the user to roughly aim the terminal at its target. Once the target enters the IR camera field-of-view (1°), our control software automatically takes over and locks onto the target. The optical terminal includes a monitoring photodiode that receives 10~\% of the return power injected in the fibre allowing fine tuning and monitoring of the link transmission. 
     
\subsection{Balloon carrier and payload}

The balloon is a 16~m$^3$ Allsopp ``Helikite'' hybrid kite/Helium balloon that can carry up to 8~kg payload including the weight of the tether. It is stabilized by light wind, but more difficult to operate in strong wind conditions. For our experiment, the tether length is set to 300~m.

Our payload is made of two parts mounted at both ends of a slim $\sim1$~m metal tube. The first is a plastic box containing the laser diode drivers connected to a USB battery mounted on the balloon itself. The second is the corner cube (2.5" diameter) and the four laser beacons at $\lambda$=1310~nm assembled on a 3D printed bloc, mounted at a right angle onto the tube and facing outward along its length. The metal tube is attached in-line with the balloon and the tether, which passes inside the tube. This arrangement ensures that the corner cube is facing the anchorage point on the ground, thereby providing a passively maintained alignment with the closeby optical terminal sufficient for the link operation, regardless of the balloon motion (translation or rotation).

\begin{figure}
\centering
\centering
\includegraphics[width=\textwidth]{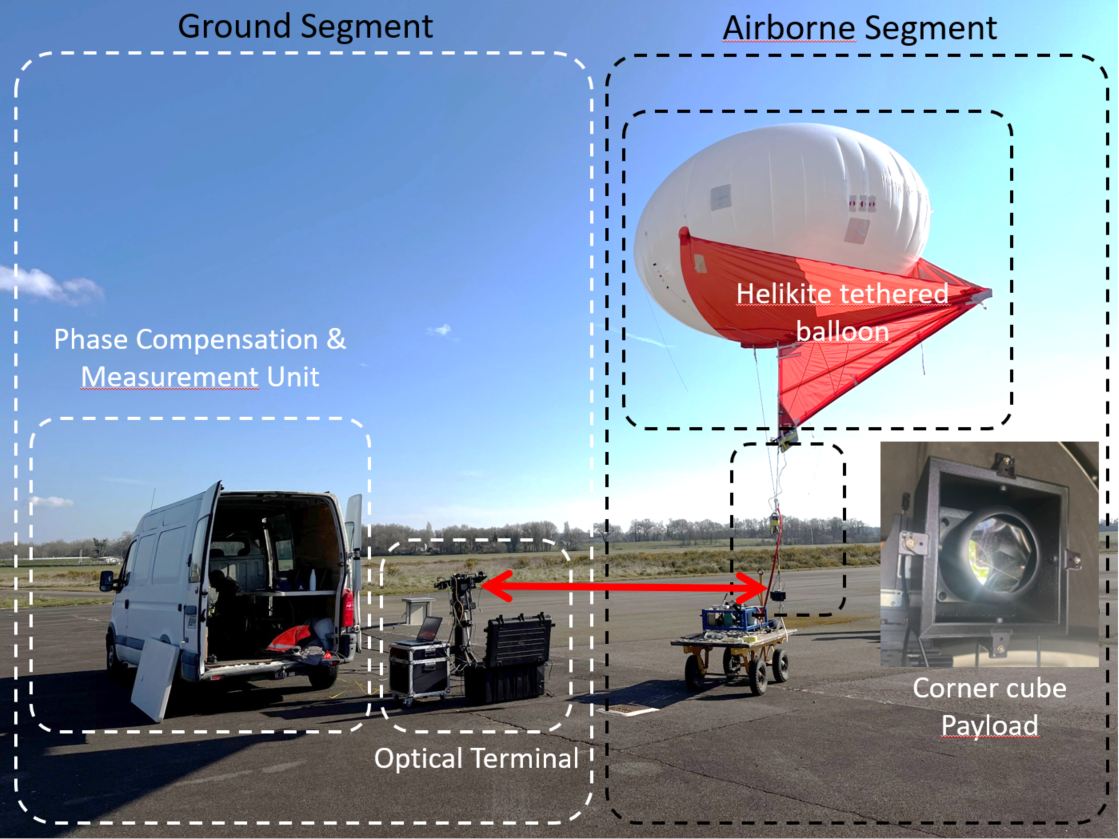}
\captionof{figure}{Picture of the experimental setup deployed at CNES in Aire sur l'Adour before balloon take-off. The inset shows the corner-cube with the four beacon diodes. The tether is anchored $\sim$~2~m from the ground terminal. Operators are located inside the van during flight.
}
\label{fig:balloon_and_payload}
\end{figure}

\section{Results} \label{sec:results}
We report here the results of a measurement campaign at the CNES balloon launch site in Aire sur l'Adour in the south-west of France in March 2023. The measurements have been achieved over 2 days of favourable weather conditions with light winds (typically $<5$~m/s) and no rain or fog that would hinder visibility. The tether length was fixed to 300~m and the light winds resulted in an almost vertical (elevation $\geq 70$ degrees) position of the balloon above the ground terminal, i.e. close to 300~m in altitude. In addition to the balloon link, a static link was established to an identical payload, installed on a mount, 300~m away from the terminal and about 1.5~m above the ground. This static link served as a reference for, and comparison to, the balloon link.

The free space-link loss i.e. the fraction of emitted power injected back into the fibre after a round-trip was about 0.025 (-16 dB) and was subject to optimizing the set-point of the tip/tilt correction, and the adjustment of the collimator and GBE focus. The latter was a compromise between maximum received power (small divergence) and minimal power fluctuations (larger divergence). For typical settings about 400~$\mu$W were launched into free space and the power injected back into the fibre was about a factor 2 above the extinction threshold of our measurement ($\sim0.5\,\mu$W).

\begin{figure}
\centering
  \centering
  \includegraphics[width=\textwidth]{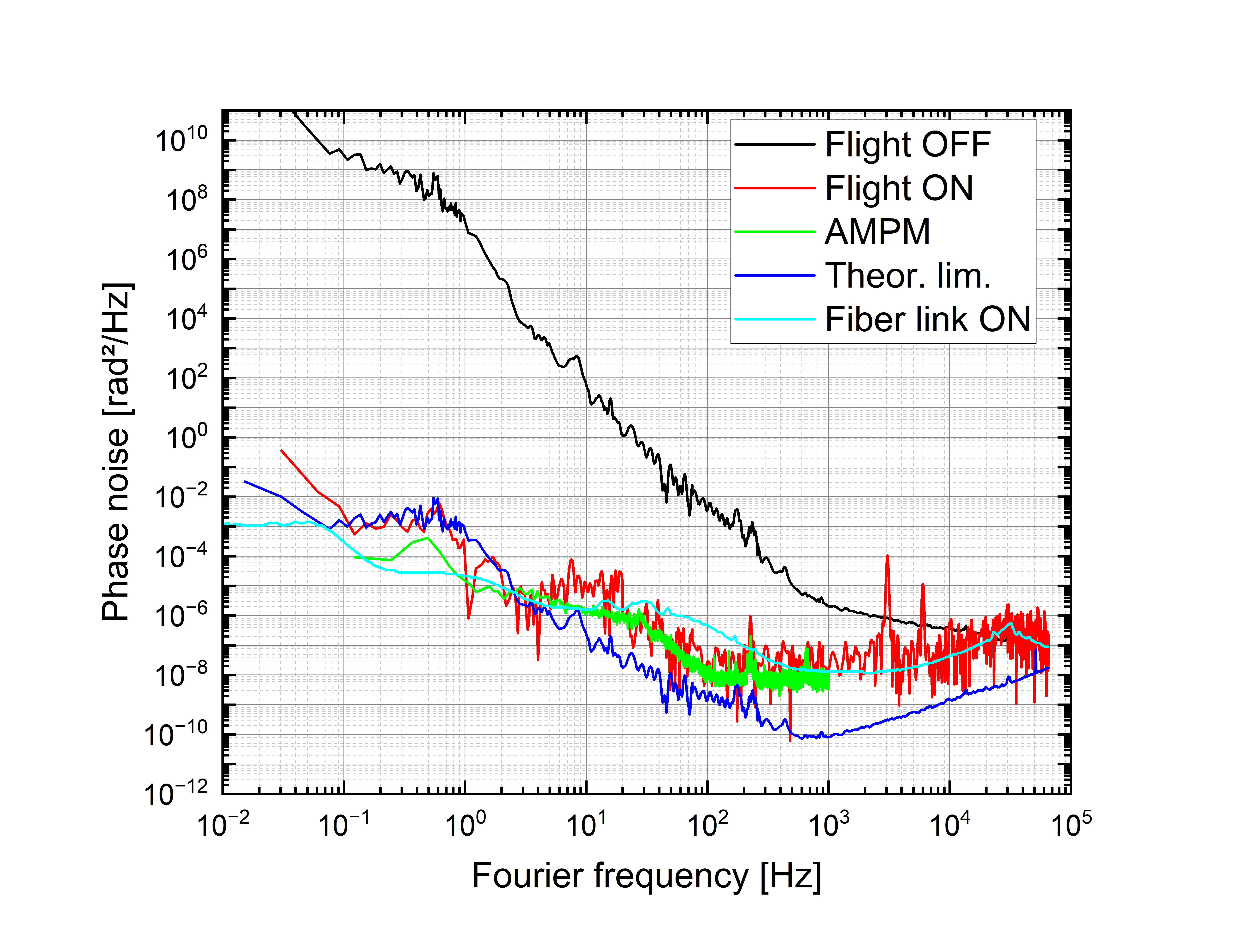}
  \captionof{figure}{In-flight phase noise measurement PSD. Black curve, free-space link uncompensated; red, free-space link compensated; green, AM/PM evaluation from monitoring photodiode measurement; dark blue, noise floor due to the link delay, evaluated from black curve; light blue, noise of a 300m compensated fibre link.}
  \label{fig:PSD_2023_vol}
\end{figure}
\begin{figure}
\centering
  \centering
  \includegraphics[width=\textwidth]{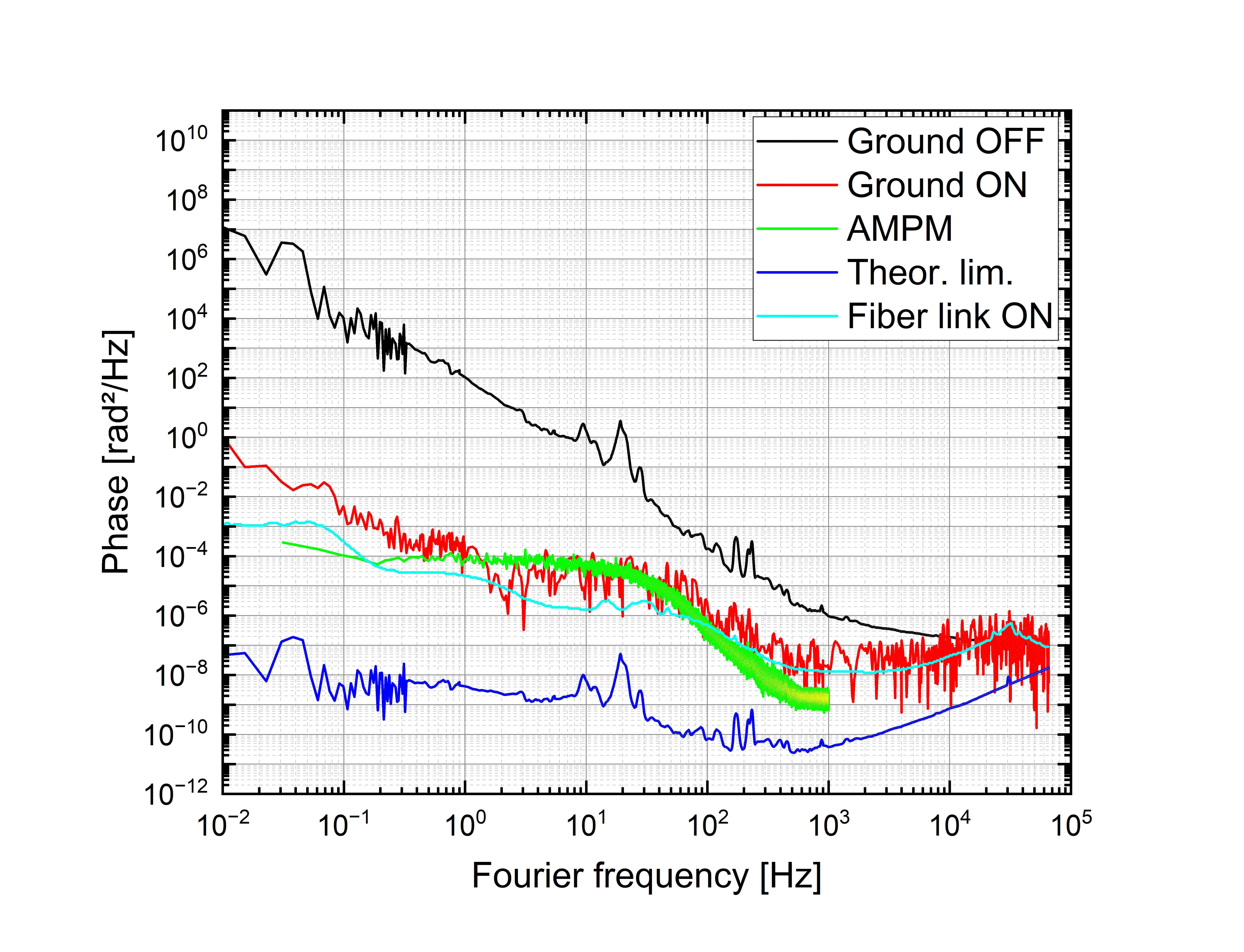}
  \captionof{figure}{On ground phase noise measurement PSD. Black curve, free-space link uncompensated; red, free-space link compensated; green, AM/PM evaluation from monitoring photo-diode measurement; dark blue, noise floor due to the link delay, evaluated from black curve; light blue, noise of a 300m compensated fibre link.}
  \label{fig:PSD_2023_sol}
\end{figure}

Figures \ref{fig:PSD_2023_vol} and \ref{fig:PSD_2023_sol} show the results in terms of phase noise PSD in both uncompensated and phase noise compensated configurations, for the balloon and ground links. All reported phase measurements are taken with the out-of-loop diode (``Meas.'' on Fig. \ref{fig:schematic_exp}). Additionally we plot the ``theoretical'' compensated performance calculated from the uncompensated data and the known link distance (see Appendix A for details), which corresponds to our measurement floor as it represents the expected performance with perfect phase noise compensation and no additional noise sources. The total duration of the measurements were 80~s (compensated in-flight), 120~s (uncompensated in-flight), 172~s (compensated ground), 505~s (uncompensated ground) \cite{footnote1}. 
The compensated and uncompensated acquisitions were carried out in quick succession to ensure similar conditions, however the on-ground and in-flight measurements were taken on different days. We also plot noise from our equivalent fibre link using the same phase noise compensation system, and the expected amplitude to phase noise (AM/PM) conversion in our out of loop measurement (see section \ref{sec:discussion}).

\begin{figure}
\centering
  \centering
  \includegraphics[width=\textwidth]{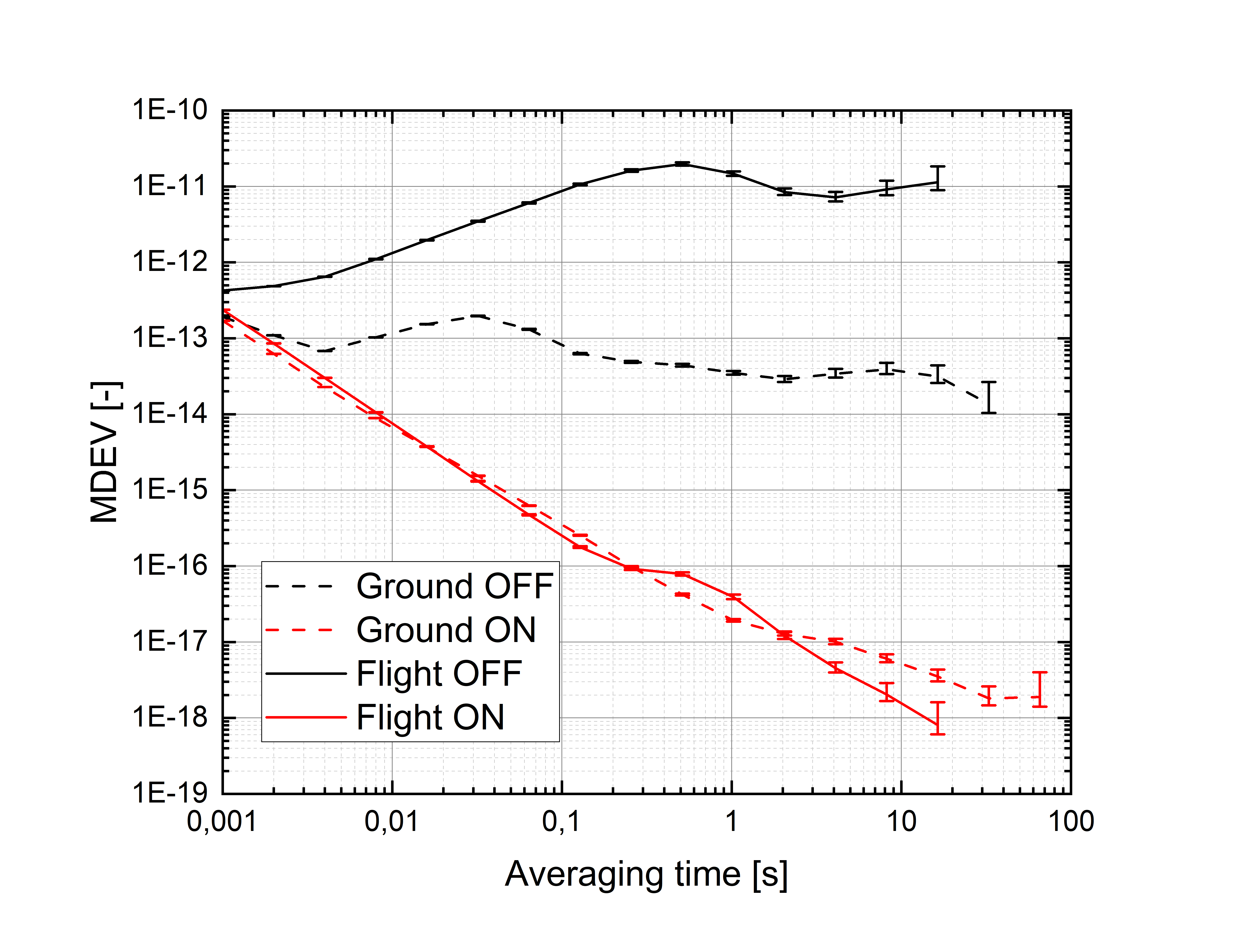}
  \captionof{figure}{Fractional frequency stability (modified Allan deviation). Same data as on figures \ref{fig:PSD_2023_vol} and \ref{fig:PSD_2023_sol}. Compensated (ON) and uncompensated (OFF) results for balloon and ground links.}
  \label{fig:MDEV}
\end{figure}

Finally, we show the frequency stability (modified Allan deviation, MDEV) in figure \ref{fig:MDEV}, demonstrating sub $10^{-18}$ performance after less than 20~s averaging. Note, that when our set-up was optimised (focal settings, PID gains, ...) no signal outages were observed even during our longest runs ($>500$~s), thus all plotted MDEV are using continuous gap-free data.

\section{Discussion} \label{sec:discussion}
We start by examining the uncompensated phase noise in flight. We note that the peak in the PSD around 0.6~Hz and corresponding bump in the MDEV around 0.5~s~\cite{footnote2} 
corresponds to oscillatory longitudinal motion of the payload, observed during all flights, including some preliminary tests in 2022. From the MDEV (or equivalently the PSD or directly from the raw phase data) we see an oscillation of the distance along the line of sight of roughly 2~mm amplitude. We also see prominent oscillations at 0.6~Hz of the applied angular corrections on the tip/tilt mirror and the astronomical mount with amplitudes of about 0.1~mrad. Taking into account the distance between the ground terminal and the tether anchor-point of about 2~m (see fig. \ref{fig:balloon_and_payload}) this accounts only partially (about 10\%) for the 2~mm longitudinal motion of the payload. We attribute the main part to a small coupling of the transverse motion of the payload (about 3~cm amplitude) into the longitudinal axis along the line of sight.

When comparing the in-flight uncompensated to ground uncompensated data (PSD or MDEV) we notice that the in-flight measurements are significantly noisier (by up to 3 orders of magnitude) at all frequencies up to 10~kHz. If the uncompensated phase measurements were dominated by instrumental noise or atmospheric turbulence we would expect the ground data to show similar or larger noise (turbulence at 1.5~m above the ground is expected to be stronger than at high elevation), but the contrary is the case. This, together with the correlation between observed angular oscillations and distance oscillations (see previous paragraph) suggests that the in-flight uncompensated phase noise is largely dominated by the motion of our balloon and payload, leading to corresponding distance changes, as one would expect.

At very high frequencies (above 10~kHz) all noise curves converge towards the noise measured on our test fibre link. In that frequency region the results are dominated by the measurement noise of our system and the phase compensation loop (DPLL) noise. The locking bump corresponding to the bandwidth of that loop is clearly visible around 30~kHz.

Turning now to the phase-noise compensated links (in flight and on ground) we first notice that in most of the frequency range we do not reach the projected performance from a purely distance limited calculation (see Appendix A). It is only reached for the in flight measurement in the region between about 0.1~Hz and 2~Hz. At lower frequencies we are limited by thermal effects on the uncompensated fibres of the experiment, in particular the out of loop measurement fibre (dashed yellow line in Fig. \ref{fig:schematic_exp}). At intermediate frequencies we are limited by amplitude to phase noise (AM/PM) conversion as discussed below. At high frequencies (above a few kHz) by the residual noise of our DPLL as seen on the fibre link.

At the lowest frequencies (below 0.1~Hz in PSD, above 10~s in MDEV) thermal effects from the uncompensated fibres are dominant. A linear fit to the phase data yields a phase drift that corresponds to a fractional frequency offset of $2\times 10^{-17}$ for the in flight data and $1\times 10^{-17}$ on ground. A quadratic fit gives a fractional frequency drift of $1.6\times 10^{-20}$/s in flight and $1.4\times 10^{-20}$/s on ground. With a thermal coefficient of the fibres of $10^{-5}$/K and for 1~m uncompensated fibre this corresponds to temperature variations of about 0.3~mK/s (=1.1~K/h), which is realistic given that the equipment was outside on the airfield, without any thermal shielding or temperature control. Part of this is expected to be due to the out of loop measurement fibre i.e. not actually part of the link. In any case, some even modest optimisation (reduce uncompensated fibre lengths, add thermal shielding and control and/or correct in post-processing) should allow reducing long term effects below $10^{-18}$ in fractional frequency offset and well below $10^{-20}$/s in fractional frequency drift.

At intermediate frequencies (a few Hz to about a kHz) our noise limitation comes from AM/PM conversion. The effect is located mainly in the mixer used to down-convert the 80~MHz out of loop measurement to 1~MHz. As such it is not part of the link itself, but only present in the verification measurement. We have checked this effect by controlled modulation of the signal amplitude on the ground link at CNES premises in Toulouse. The signal amplitude was modulated by sinusoidally varying the tip/tilt mirror. The observed phase modulation was proportional to the generated amplitude modulation with a coefficient \cite{footnote3} 
of $K_{AP} \approx 0.09$ rad. Using that coefficient we converted the amplitude fluctuations measured on our power monitoring diode (see Fig. \ref{fig:schematic_exp}) to phase fluctuations, as plotted on figures \ref{fig:PSD_2023_vol} and \ref{fig:PSD_2023_sol}. They fit well to the measured phase noise. This measurement artefact can easily be avoided in the future by redesigning our measurement chain. We think that it is not a fundamental limitation of the link.

Finally, we discuss the possible extension to larger distances. Indeed, applications like chronometric geodesy will require links to higher altitudes (1-10~km) leading to line of sight distances of tens of km. Our current configuration was designed for technology demonstration (tracking, phase noise compensation) and is certainly not suited for such large distances. However, the obtained results indicate that with appropriate design (10~cm size optics, improved link efficiency, higher laser power, \dots)  distances of several tens of km should be possible. If necessary, the constraints on the power budget for such long distances could be reduced by using an on-board bidirectional amplifier (EDFA on Fig. \ref{fig:schematic_exp}) provided the mass and power budget of the airborne carrier allows it. In the shorter term, we plan to study and optimize the one-way (ground-balloon) power budget of our short link in order to better understand the requirements for long distance links.

\section{Outlook and conclusion} \label{sec:conclusion}

In conclusion, we presented the characterization of a folded free-space optical link via a corner cube onboard a tethered balloon. The results, obtained during a flight campaign in March 2023, demonstrate a fractional frequency stability of $8\times 10^{-19}$ after 16~s averaging time, which is the best result for airborne experiments reported so far. In addition, we extensively identified and discussed the main limiting noise sources and how they could be overcome in future implementations.

This first step paves the way to point-to-point ground links via an airborne balloon relay (type (i) link mentioned in the introduction). This will be achieved through a miniaturization of our active system to fit within the balloon capability. Particular attention will be paid to the improvement of the pointing and stabilization performance which appears to be limiting in case of windy weather conditions. Such free-space comparisons will be useful for many applications, like chronometric geodesy, using operational and transportable in-field systems via airborne relays.

\section*{Acknowledgements}

This work was conducted in the framework of the Ph.D. thesis of N.M. co-funded by the Labex First-TF and CNES, and the Ph.D. thesis of S.F. co-funded by AID and CNES.

The authors would like thank the CNES balloon aeronauts (Brice Bellanger) for their time and expertise, and colleagues from Météo-France (Axel Roy) for feedback and data concerning their Helikite experience. We also thank Jean-Daniel Deschênes for help with the DPLL software. Last but not least we acknowledge fruitful collaboration and discussion over the last years with our colleagues from the University of Western Australia (Sascha Schediwy, David Gozzard, Benjamin Dix-Matthews among others), it is great and much easier to advance together! 

\section*{Disclosures}

The authors declare no conflicts of interest.

\section*{Data availability}

Data underlying the results presented in this paper are not publicly available at this time but may be obtained from the authors upon reasonable request.

\section*{Appendix A: Calculation of the theoretical limit for the compensated noise}
As discussed in section \ref{sec:discussion} the noise of the uncompensated link is dominated by the balloon motion. We model it by a virtual frequency shift $\delta\nu(t)$ that occurs exactly half way between the Tx and Rx terminals. The received frequency in the uncompensated link is
\begin{equation}
\nu_{Rx}^{OFF}(t)=\nu_L + \delta\nu (t-T/2),
\end{equation}
where $\nu_L$ is the laser frequency and $T$ is the flight time between Tx and Rx (in our case $T=2L/c$, where $L=300$~m). Thus we have for the PSD of the out of loop diode (that measures $\nu_{Rx}^{OFF}(t)-\nu_L$): $S_{OFF}(f) = S_{\delta\nu}(f)$.

For the compensated link the frequency at the Rx end is
\begin{equation}\label{equ:nuON}
\nu_{Rx}^{ON}(t)=\nu_L + \Delta\nu(t-T) + \delta\nu(t-T/2),
\end{equation} 
where $\Delta\nu(t)$ is the frequency correction applied by the AOM at the Tx end. We neglect the Rx-AOM and the constant frequency of the Tx-AOM, which are irrelevant for this calculation.

The error signal at the Tx end is set to zero by the DPLL hence
\begin{eqnarray}\label{equ:err_Dix}
0 &=& \nu_L-\nu_{ret}(t) \nonumber \\
&=& \nu_L - \big(\nu_L+\Delta\nu(t-2T) \nonumber \\
&+& \delta\nu(t-3T/2)+\delta\nu(t-T/2)+\Delta\nu(t)\big) \,,
\end{eqnarray}
where $\nu_{ret}(t)$ is the frequency of the signal after a full Tx-Rx-Tx round trip. A first order expansion leads to
\begin{eqnarray}\label{equ:Dnu}
2\Delta\nu(t) - 2T\Delta\dot{\nu}(t) &\simeq& -(2\delta\nu(t) - 2T\delta\dot{\nu}(t)) \nonumber \\
\implies \,\, \Delta\nu(t) &\simeq& -\delta\nu(t) \,.
\end{eqnarray}

Substituting (\ref{equ:Dnu}) into (\ref{equ:nuON}) gives
\begin{equation}\label{equ:nuON_corr}
\nu_{Rx}^{ON}(t)=\nu_L - \delta\nu(t-T) + \delta\nu(t-T/2) \,.
\end{equation}

Finally, the out of loop diode measures is $\nu_{Rx}^{ON}(t)-\nu_L$, and the corresponding PSD is \cite{footnote4} 
\begin{eqnarray}\label{equ:S_ON}
S_{ON}(f) &=& 2(1-\cos(\pi f T))S_{\delta\nu}(f) \nonumber \\
&=& 2(1-\cos(\pi f T))S_{OFF}(f) \,.
\end{eqnarray}

\bibliography{TOFU}

\begin{thebibliography}{10}
\bibitem{footnote1}
All data acquisitions were intentionally stopped at durations sufficient for our purposes ($\sim 10^2$~s, beyond that we only see thermal effects, see sect. \ref{sec:discussion}) in order to avoid gaps, although none were observed, and to leave time for different tests and variations of settings.
\bibitem{footnote3}
We define the AM/PM coefficient by $\Delta\phi(t)=K_{AP}\frac{\Delta P(t)}{P_0}$, where $\Delta\phi(t)$ and $\Delta P(t)$ are the phase and power variations and $P_0$ is the average power.
\end{thebibliography}

\end{document}